
\overfullrule 0pt

\pageno=1

\count30=0                      
\count31=0                      
\count32=0                      
\count33=0                      

\def\chapno{\global\count31=0
            \global\count32=0
            \global\count33=0
            \global\advance\count30 by1
            \number\count30}

\def\secno{\global\advance\count31 by1
           \number\count30 .\number\count31}

\def\forno{\eqno(\global\advance\count32 by1
            \number\count30 .\number\count32)}

\def\prono{\global\advance\count33 by1
            \number\count30 .\number\count33}

\long\def\proof#1{\noindent{\bf{Proof.}}\quad#1\qedd}
\def\qedd{\hfill\hbox{\vrule\vbox{\hrule\kern2truept\hbox{\kern2truept
\vbox to2pt{\hsize=2pt}\kern2truept}\kern2truept\hrule}\vrule}\vskip .5cm}

\def\oi{o_{\infty }}
\def\gli{gl_{\infty }}
\def\glio{\overline {gl_{\infty }}}
\def\oio{\overline {o_{\infty }}}
\def\Ai{A_{\infty}}
\def\Bi{B_{\infty}}

\vskip .5cm
\centerline{\bf THE $n$--TH REDUCED BKP HIERARCHY, THE STRING}
\vskip .1cm
\centerline{\bf EQUATION AND
$BW_{1+\infty}$--CONSTRAINTS}
\vskip 1cm
\centerline {JOHAN VAN DE LEUR\footnote {*}{The research of Johan van de Leur
is
financially supported by the ``Stichting Fundamenteel Onderzoek der
Materie (F.O.M.)''. E-mail: vdleur@math.utwente.nl}}
\vskip 2cm
\centerline {\bf Abstract}
\midinsert\narrower\narrower\narrower\narrower
\noindent We study the BKP hierarchy and its $n$--reduction, for the case that
$n$ is odd. This is related to the principal realization of the basic module
of the twisted affine Lie algebra $\hat{sl}_n^{(2)}$.
We show that the following two statements  for a BKP $\tau$ function are
equivalent: (1) $\tau$ is  is $n$--reduced and  satisfies the string equation,
i.e. $L_{-1}\tau=0$, where $L_{-1}$ is an element of some `natural'
Virasoro algebra. (2)  $\tau$ satisfies the vacuum constraints of the
$BW_{1+\infty}$ algebra.  Here $BW_{1+\infty}$ is the natural analog of the
$W_{1+\infty}$ algebra, which plays a role in the KP case.
\endinsert
\vskip .5cm

\beginsection \chapno . Introduction\par
\secno .
In recent years KdV type hierarchies have been related to 2D gravity. To be
slightly more precise  (see [Dij] for the details and references),
the square root of the partition function of the Hermitian $(n-1)$--matrix
model
in the continuum limit is the $\tau$--function of the $n$--reduced
Kadomtsev Petviashvili (KP) hierarchy. The partition function is then
characterized by the so-called string
equation:
$$L_{-1}\tau={1\over n}{\partial \tau\over\partial x_1},$$
where $L_{-1}$ is an element of  the $c=n$  Virasoro algebra,
wich is related to the principal realization of the affine lie algebra
$\hat{sl}_n$,
or rather $\hat{gl}_n$. Let $\alpha_k=-kx_{-k},\ 0,\ {\partial\over\partial
x_k}$ for
$k<0,\ k=0,\ k>0$, respectively, then
$$L_k={1\over 2n}\sum_{\ell\in {\bf Z}}:\alpha_{-\ell}\alpha_{\ell+nk}:
+\delta_{0k}{n^2-1\over 24n}.
\forno
$$

By making the shift $x_{n+1}\mapsto x_{n+1}+{n\over n+1}$, we modify the
origin of the $\tau$--function and thus obtain the following
form of the string equation:
$$L_{-1}\tau=0.\forno$$

Actually, it can be shown ([FKN], [G] and [AV]) that the above conditions,
$n$--th reduced KP and equation (1.2) (which from now on we will call
the string equation),
on a $\tau$--function of the KP hierarchy imply
more general constraints, viz. the vacuum constraints of the $W_{1+\infty}$
algebra. This last condition is reduced to the vacuum conditions of the $W_n$
algebra when some redundant variables are eliminated.

The $W_{1+\infty}$ algebra is the central extension of the Lie algebra of
differential
operators on ${\bf C}^{\times}$.
This central extension was discovered by Kac and Peterson in 1981 [KP] (see
also [R], [KR]). It has as basis the operators
$W^{(\ell+1)}_k=-s^{k+\ell}({\partial\over\partial s})^{\ell}$, $\ell\in{\bf
Z}_+$,
$k\in{\bf Z}$, together with the central element $c$. There is a well-known way
how to express these elements in the elements of the Heisenberg algebra,
the $\alpha_k$'s. The $W_{1+\infty}$ constraints then are
$$\hat W^{(\ell+1)}_k\tau=\{W^{(\ell+1)}_k +\delta_{k,0}c_{\ell+1}\}\tau=0
\ \hbox{ \rm for } -k\le \ell\ \hbox{\rm and}\ \ell\ge 0.
$$
For the above $\tau$--function, $\hat W^{(1)}_k=-\alpha_{nk}$ and
$\hat W^{(2)}_k=L_k-{nk+1\over n}\alpha_{nk}$.

\noindent \secno.
In this paper we study the $n$--th reduced BKP hierarchy, where we assume that
$n$
is odd. This reduction is related to the principal realization of the
basic module of the affine Lie algebra
$\hat{sl}_n^{(2)}$. A $\tau$--function of the $n$--th reduced BKP hierarchy
is a function in the variables $x_1,\ x_3,\ x_5,\ldots$ with the
restriction that $\tau$ is independent of the variables
$x_{jn}$ for $j=1,3,5, \ldots$.  For the principal realization of the
basic module of this affine Lie algebra
$\hat{sl}_n^{(2)}$, there exists a `natural' Virasoro algebra. Now assuming
that
this $\tau$--function also satisfies $L_{-1}\tau=0$, we show that $\tau$
also satisfies the vacuum constraints of the $BW_{1+\infty}$ algebra.
The best way to describe    $BW_{1+\infty}$ is as a subalgebra of
 $W_{1+\infty}$. Let $\iota$ be a  linear anti--involution on
 $W_{1+\infty}$ defined by:
$$\iota (s)=-s,\quad \iota (s{\partial \over \partial s})=-s{\partial \over
\partial s}\quad\hbox{\rm and}\ \iota(c)=-c,\forno$$
then
$$BW_{1+\infty}=\{w\in W_{1+\infty}|\iota(w)=-w\}.\forno$$
Let
$$W_{j\over2}^{(k+1)}=
-s(s^{{j}+2k}
({\partial\over \partial s^2})^k-
(-)^{j+k}({\partial\over \partial s^2})^k
s^{{j}+2k})s^{-{1}},$$  we  then show that
$$\{W_{j\over2}^{(k+1)}+\delta_{j0}c_{k+1}\}\tau =0\quad \hbox{\rm for}\
j\ge -2k\ \hbox{\rm and}\ k\le 0,
$$
here $c_{k+1}$ are constants that depend on $n$.

Many of the results presented in the sections 2--4 are well--known
and can be found in e.g. [DJKM], [Sh2] and [Y].

Finally , it is a pleasure to thank Frits Beukers for useful discussions
and the Mathematical Institute of the University of Utrecht for
computer and e-mail facilities.

\beginsection \chapno .  The spin representation of $\oi$, $\Bi$
and the BKP hierarchy in the fermionic picture\par
\secno .
Let $\glio$ be the Lie algebra of complex infinite dimensional matrices
such that all nonzero entries are within a finite distance from the
main diagonal, i.e.,
$$\glio =\{(a_{ij})_{i,j\in {\bf Z}}| g_{ij}=0 \ {\rm if}\ |i-j|>>0\}.$$
The elements $E_{ij}$, the matrix with the $(i,j)$--th entry 1 and 0 elsewhere,
for $i,j\in{\bf Z}$ form a basis of a subalgebra
$\gli\subset\glio$.
The Lie algebra $\glio$ has a universal central extension
$\Ai=\glio\oplus{\bf C}c_A$ with the Lie bracket defined by
$$[a+\alpha c_A, b+\beta c_A]=ab-ba+\mu(a,b)c_A \forno$$
for $a,b\in\glio$ and $\alpha,\beta\in{\bf C}$;
here $\mu$ is the following 2--cocycle:
$$\mu(E_{ij},E_{kl})=\delta_{il}\delta_{jk}(\theta(i)-\theta(j)),\forno$$
where the function $\theta:{\bf R}\to{\bf R}$ is defined by
$$\theta(i)=\cases{0&if $i>0$,\cr 1&if $i\le 0$,\cr} \forno$$
\secno .
Define on $\glio$ the following linear anti--involution:
$$\iota(E_{jk})=(-)^{j+k}E_{-k,-j} \forno$$
Using this anti--involution we define the Lie algebra $\oio$ as
a subalgebra of $\glio$:
$$\oio=\{a\in\glio |\iota(a)=-a\} \forno$$
The elements $F_{jk}=E_{-j,k}-(-)^{j+k}E_{-k,j}=-(-)^{j+k}F_{kj}$
with $j<k$ form a basis of $\oi=\oio\cap\gli$. The 2--cocycle $\mu$
on $\glio$ induces a 2--cocycle on $\oio$, and hence we can define
a central extension $\Bi=\oio\oplus{\bf C}c_B$ of $\oio$, with Lie
bracket
$$[a+\alpha c_B,b+\beta c_B]=ab-ba+{1\over 2}\mu(a,b)c_B \forno$$
for $a,b\in\oio$ and $\alpha,\beta\in{\bf C}$.

\noindent\secno .
We now want to consider highest weight representations of $\oi$ and
$\Bi$. For this purpose we introduce the Clifford algebra BCl as the
associative algebra on the generators $\phi_j$, $j\in{\bf Z}$, called
{\it neutral free fermions}, with defining relations
$$\phi_i\phi_j+\phi_j\phi_i=(-)^i\delta_{i,-j}. \forno$$
We define the spin module $V$ over BCl as the irreducible module with
highest weight vector the {\it vacuum vector} $|0>$ satisfying
$$\phi_j|0>=0\quad\hbox{ \rm for}\ j>0. \forno$$
The elements $\phi_{j_1}\phi_{j_2}\cdots\phi_{j_p}|0>$
with $j_1<j_2<\cdots <j_p\le 0$ form a basis of $V$. Then
$$\eqalign{\pi(F_{jk})&={(-)^j\over 2}(\phi_j\phi_k-\phi_k\phi_j),\cr
     \hat{\pi}(F_{jk})&= (-)^j :\phi_j\phi_k:,\cr
     \hat{\pi}(c_B)&=I,\cr}
\forno$$
where the normal ordered product $:\quad :$
is defined as follows
$$:\phi_j\phi_k :=
\cases{\phi_j\phi_k& if $k>j$,\cr
       {1\over2}(\phi_j\phi_k-\phi_k\phi_j)& if $j=k$,\cr
       -\phi_k\phi_j& if $k<j$,\cr}
\forno
$$
define representations of $\oi$, respectively $\Bi$.

When restricted to $\oi$ and $\Bi$, the spin module $V$ breaks
into the direct sum of two irreducible modules.
To describe this decomposition we define a ${\bf Z}_2$--gradation
on $ V$ by introducing a chirality operator $\chi$ satisfying
$\chi|0>=|0>,\ \chi\phi_j+\phi_j\chi=0$ for all $j\in{\bf Z}$,
then
$$V=\bigoplus_{\alpha\in {\bf Z}_2} V_{\alpha}\quad\hbox{\rm where}\
V_{\alpha}=\{v\in V|\chi v=(-)^{\alpha}v\}.
$$
Each module $V_{\alpha}$ is an irreducible highest weight
module with highest weight vector $|0>$, $|1>=\sqrt 2\phi_0|0>$
for $V_0,\ V_1$, respectively, in the sense that
$$\eqalign{\pi(F_{-i,j})|\alpha>&=\hat{\pi}(F_{-i,j})|\alpha>=0\quad
                             \hbox{\rm for}\ i<j,\cr
         \pi(F_{-i,i})&=-{(-)^i\over 2}|\alpha>\quad \hbox{\rm for}\ i>0,\cr
         \hat{\pi}(F_{-i,i})&=0.\cr}
\forno$$
\secno .
Now we define the operator $Q$ on $V$ by
$$\eqalign{Q|0>&=\sqrt 2\phi_0|0>,\cr
         Q\phi_j&=\phi_jQ\quad\hbox{\rm for all}\ j\in{\bf Z}.\cr}
\forno$$
Clearly $Q^2=I$. Let $S$ be the following operator on $V\otimes V$:
$$S=\sum_{j\in{\bf Z}} (-)^j \phi_j\otimes \phi_{-j}.\forno$$
Then
$$S(|0>\otimes|0>)=\phi_0|0>\otimes \phi_0|0>
                 ={1\over 2}Q|0>\otimes Q|0>.$$
Notice that both $Q$ and $S$ commute with the action of $\oi$.
Let $\tau\in V_0$, then we define the BKP equation (in the fermionic picture)
to be the following equation:
$$S(\tau\otimes\tau)={1\over 2}Q\tau\otimes Q\tau. \forno$$
One can show [H] that there exists a group $G$ for which $\tau$
an element of the group orbit of the vacuum vector $|0>$ is, if
and only if $\tau$ satisfies (2.14). But since we will not use the
group in the rest of this paper, we will not prove this statement
here.

\beginsection \chapno . Vertex operators and the BKP hierarchy
in the bosonic picture \par
\secno .
Define the following two generating series (fermionic fields):
$$\phi^{\pm}(z)=\sum_{j\in{\bf Z}} \phi_j^{\pm} z^{-{j\over 2}-{1\over 2}}
               =\sum_{j\in{\bf Z}}(\pm)^j\phi_j z^{-{j\over 2}-{1\over 2}}.
$$
Using this we define
$$\alpha (z)=\sum_{j\in {1\over 2}+{\bf Z}}\alpha_j z^{-j-1}
            ={1\over 2}:\phi^+(z)\phi^-(z):,
\forno$$
then one has (see e.g. [tKL] for details):
$$\eqalign{[\alpha_j,\phi^{\pm}(z)]&=\pm z^j\phi^{\pm}(z),\cr
         [\alpha_j,\alpha_k]&=j\delta_{j,-k}\cr}
$$
and
$$\phi^{\pm}(z)={Q\over\sqrt 2}\exp (\mp\sum_{j<0}{z^{-j}\over j}\alpha_j )
                               \exp (\mp\sum_{j>0}{z^{-j}\over j}\alpha_j ).
\forno$$
Then it is straightforward that one has the following isomorphism (see [tKL]):
$\sigma :V\to  {\bf C}[\theta,x_1,x_3,\cdots]$, where $\theta^2=0$,
$x_ix_j=x_jx_i$, $\theta x_j=x_j\theta$ and
$V_{\alpha}=\theta^{\alpha}{\bf C}[x_1,x_3,\cdots].$
Now $\sigma(|0>)=1$ and
$$\eqalign{\sigma\alpha_j\sigma^{-1}&=\cases{-jx_{2j}& if $j<0$,\cr
                                          {\partial\over\partial x_{2j}}&
                                           if $j>0$,\cr}\cr
         \sigma Q\sigma^{-1}&=\theta+{\partial\over\partial\theta}.\cr}\forno
$$
Hence
$$\sigma\phi^{\pm}(z)\sigma^{-1}=
{\theta+{\partial\over\partial\theta}\over\sqrt 2}z^{-{1\over 2}}
\exp(\pm\sum_{j>0,{\rm odd}}x_jz^{j\over2})
\exp(\mp2\sum_{j>0,{\rm odd}}{\partial\over\partial x_j}{z^{-{j\over 2}}\over
j}).
\forno$$
\secno .
We first rewrite the BKP hierarchy (2.14):
$${\rm Res}_{z=0} dz \phi^+(z)\tau\otimes\phi^-(z)\tau=
{1\over 2}Q\tau\otimes Q\tau.
\forno$$
Here ${\rm Res}_{z=0} dz \sum_j f_jz^j=f_{-1}$. Now replace $z$ by $z^2$
and use (3.4), then (3.5) is equivalent to
$${\rm Res}_{z=0}{dz\over z}
\exp\sum_{j>0,{\rm odd}}x_jz^{j}
\exp (-2\sum_{j>0,{\rm odd}}{\partial\over\partial x_j}{z^{-{j}}\over j})
\tau\otimes
\exp -\sum_{j>0,{\rm odd}}x_jz^{j}
\exp (2\sum_{j>0,{\rm odd}}{\partial\over\partial x_j}{z^{-{j}}\over j})
\tau=\tau\otimes\tau.
\forno$$
Equation (3.6) is called the BKP hierarchy in the bosonic picture.
It is straightforward, using change of variables and Taylor's
formula, to rewrite (3.6) into a generating series of Hirota bilinear equations
(see e.g. [DJKM]).

\beginsection \chapno . The BKP hierarchy in terms of formal
pseudo--differential operators \par
\secno .
We start by reviewing some of the basic theory of formal pseudo--differential
operators (see e.g. [DJKM], [Sh1] and [KL]).
We shall work over the algebra ${\it A}$ of formal power
series over ${\bf C}$ in indeterminates $x = (x^{k})$, where
$k = 1,3,5,\ldots $.  The indeterminate
$x_{1}$ will be viewed as variables and
$x_k$ with $k \ge 3$ as parameters.  Let
$\partial = {\partial\over\partial x_{1}},$
{\it a formal matrix pseudo-differential
operator} is an expression of the form
$$P(x,\partial ) = \sum_{j \leq N} P_{j}(x)\partial^{j}, \forno$$
where $P_{j}\in{\it A}$.   Let $\Psi$ denote the vector
space over ${\bf C}$ of all expressions (4.1).  We have a linear
isomorphism $S :\ \Psi \to {\it A}((z))$ given by
$S(P(x,\partial )) = P(x,z)$.  The series
$P(x,z)$ in indeterminates $x$ and $z$ is called the {\it symbol} of
$P(x,\partial )$.

Now we may define a product $\circ$ on $\Psi$ making it an associative
algebra:
$$S(P \circ Q) = \sum^{\infty}_{n = 0} {1\over n!}
{\partial^{n}S(P)\over\partial z^{n}} \partial^{n}S(Q).\forno$$
{}From now on, we shall drop the multiplication sign $\circ$ when no ambiguity
may arise.
One defines the differential part of $P(x,\partial )$ by
$P_{+}(x,\partial ) = \sum^{N}_{j=0} P_{j}(x) \partial^{j},$
and let $P_{-} = P-P_{+}$.  We have the corresponding vector space
decomposition:
$$\Psi = \Psi_{-} \oplus \Psi_{+}. \forno$$
One defines a linear map $*: \Psi \rightarrow \Psi$ by the
following formula:
$$(\sum_{j} P_{j}\partial^{j})^{*} = \sum_{j} (-\partial )^{j}
\circ  P_{j}. \forno$$
Note that $*$ is an anti-involution of the algebra
$\Psi$.
There exists yet another anti--involution, viz. (see also [Sh2])
$$\iota^* P=\partial^{-1}P^*\partial\forno$$

Introduce the following notation
$$z \cdot x = \sum^{\infty}_{k=1} x_{2k-1} z^{2k-1}.$$
The algebra $\Psi$ acts on the space $U_{+}$ (resp.
$U_{-}$) of formal oscillating matrix functions of the form
$$\sum_{j \leq N} P_{j}z^{j}e^{z\cdot x}\ \ \ (\hbox{\rm resp.}\ \sum_{j
\leq N} P_{j}z^{j} e^{-z\cdot x}), \ \hbox{\rm where}\ \ P_{j} \in \ {\it A}
,$$
in the obvious way:
$$P(x)\partial^{j}e^{\pm z\cdot x} = P(x)(\pm z)^{j} e^{\pm z\cdot x}.$$
One has the following fundamental lemma (see [DJKM],[[K],[KL] or [Sh1]).

\proclaim Lemma \prono . If $P,Q \in \Psi$ are such that
$${\rm Res}_{z=0} dz P(x,\partial) e^{z\cdot x}
Q(x^{\prime},\partial^{\prime}) e^{-z\cdot x^{\prime}} = 0,
\forno$$
then $(P \circ Q^{*})_{-} = 0$.
\par
\noindent\secno .
Divide (3.6) by $\tau$, remove the tensor symbol $\otimes$
and write $x$, respectively $x'$ for the first, respectively the second,
term of the tensor product, then (3.6) is equivalent to
$${\rm Res}_{z=0} {dz\over z}w(x,z)w(x',-z)=1,
\forno$$
where $w(x,z)=P(x,z)e^{x\cdot z}=\sum_{i\ge 0}P_iz^{-i}e^{x\cdot z}$ and
$$\eqalign{P(x,z)
&={\exp (-2\sum_{j>0}{\partial\over\partial x_j}{z^{-j}\over j})\tau(x)
   \over\tau(x)}\cr
&={\tau(x_1-{2\over z},x_3-{2\over 3z^3},\cdots)\over \tau(x)}
 =:{\tilde\tau(x,z)\over\tau(x)}.\cr}
\forno$$
Notice that $P_0=1$. Now differentiate (4.7) to $x_k$, then we obtain
$${\rm Res}_{z=0}{dz\over z}({\partial P(x,z)\over\partial x_k}+P(x,z)z^k)
e^{x\cdot z}P(x',-z)e^{-x'\cdot z}=0.
\forno$$
Now using lemma 4.1 we deduce that
$$(({\partial P\over\partial x_k}+P\partial^k)\partial^{-1}P^*)_-=0.$$
{}From the case $k=1$ we then deduce that $P^*=\partial P^{-1}\partial^{-1}$
or
$$P^{-1}=\iota^*(P), \forno$$
if $k\ne 1$, one thus obtains
$${\partial P\over\partial x_k}= -(P\partial^kP^{-1}\partial^{-1})_-\partial P.
\forno$$
Since $k$ is odd, $\iota^*(P\partial^kP^{-1})=-P\partial^kP^{-1}$ and hence
$(P\partial^kP^{-1}\partial^{-1})_-\partial=(P\partial^kP^{-1})_-$. So (4.11)
turns into Sato's equation:
$${\partial P\over \partial x_k}=-(P\partial^kP^{-1})_-P.\forno$$
\secno .
Define the operators
$$L=P\partial P^{-1},\quad \Gamma=\sum_{j>0} jx_j\partial^{j-1}\quad\hbox{\rm
and}
\quad M=P\Gamma P^{-1}.\forno$$
Then $[L,M]=1$ and $\iota^*(L)=-L$. Let $B_k=(L^k)_+$, using (4.12) one deduces
the following Lax equations:
$$\eqalign{{\partial L\over \partial x_k}&=[B_k,L],\cr
         {\partial M\over\partial x_k}&=[B_k,M].\cr}
\forno$$
The first equation of (4.14) is equivalent to the following Zakharov Shabat
equation:
$${\partial B_j\over\partial x_k}-{\partial B_k\over\partial x_j}=[B_k,B_j],
\forno$$
which are the compatibility conditions of the following linear problem for
$w=w(x,z)$:
$$Lw=zw, \quad Mw={\partial w\over\partial z}\quad\hbox{\rm and}\quad
{\partial w\over\partial x_k}=B_kw.\forno$$
\secno .
The formal adjoint of the wave function $w$ is (see [DJKM]):
$$\eqalign{w^*=w^*(x,z)&=P^{*-1}e^{-x\cdot z}\cr
                     &=\partial P\partial^{-1}e^{-x\cdot z}.\cr}\forno$$
Now $L^*=-\partial L\partial^{-1}=-\partial P\partial P^{-1}\partial^{-1}$
and $M^*=\partial P\partial^{-1}\Gamma\partial P^{-1}\partial^{-1}$, so
$[L^*,M^*]=-1$ and
$$L^*w^*=zw^*, \quad M^*w^*=-{\partial w^*\over\partial z}\quad\hbox{\rm
and}\quad
{\partial w^*\over\partial x_k}=-(L^{*k})_+w^*=-B^*_kw^*.\forno$$
Finally, notice that by differentiating the bilinear identity (4.7) to $x_1'$
we obtain
$${\rm Res}_{z=0}dz w(x,z)w^*(x',z)=0.\forno$$
\beginsection \chapno . The $n$--th reduced BKP hierarchy\par
\secno .
{}From now on we assume that $n$ is an odd integer.
Let $\omega=e^{2\pi i\over n}$, then it is well--known [DJKM], [tKL] that the
fields
$$A_j(z)=:\phi^+(z)\phi^-(\omega^{2j}z):\quad\hbox{for}\ j=1,2,\ldots,n
\forno$$
generate the principal realization of the basic representation of the
Lie algebra $\hat{gl}_n^{(2)}$.
Using (3.4), one can express the fields (5.1) for $j\ne n$ in terms of the
$x_k$ and
$\partial\over\partial x_k$'s. These fields for $j\ne n$ are independent of
$x_{kn}$ and $\partial\over \partial x_{kn}$. Hence in order to describe the
representation theory of $\hat{sl}_n^{(2)}$ one only has to remove
$x_{kn}$ and $\partial\over \partial x_{kn}$
in $A_n(z)=2\alpha(z)$.

\noindent\secno .
The reduction of the BKP hierarchy to $\hat{sl}_n^{(2)}$,
considered in the previous subsection, is called the $n$--th reduced
BKP hierarchy. Hence, from now on we will call a BKP
$\tau$--function $n$--reduced, if it satisfies
$${\partial\tau\over\partial x_{kn}}=0\quad\hbox{for}\ k=1,3,5,\ldots
\forno$$
Using Sato's equation (4.12) this implies the following two equivalent
conditions:
$$\eqalign{{\partial w\over\partial x_{kn}}&=z^{kn}w,\cr
           (L^{kn})_-&=0\qquad\hbox{for}\ k=1,3,5,\ldots\cr}
$$
Hence $L^n$ is a differential operator.

\beginsection \chapno . The string equation \par
\secno .
The principal realization of $\hat{gl}_n^{(2)}$ has a natural
Virasoro algebra. In [tKL] it was shown that the following two sets of
operators have the same action on $V$ ($k\in {\bf Z}$):
$$\eqalign{L_k&={1\over 2n}\sum_{j\in {1\over 2}+{\bf Z}}
                :\alpha_{-j}\alpha_{j+nk}:
                +\delta_{k,0}({1\over 16n}+{n^2-1\over 24n}),\cr
           H_k&=\sum_{j\in{\bf Z}}{j\over 4n}
                :\phi^+_{-j}\phi^-_{j+2kn}:
                +\delta_{k,0}({1\over 16n}+{n^2-1\over 24n}).\cr}
\forno$$
So $L_k=H_k$ and
$$\eqalign{[L_k,\phi_j^{\pm}]&=-({j\over 2n}+{k\over2})\phi^{\pm}_{j+2kn},\cr
           [L_k,L_j]&=(k-j)L_{k+j}+\delta_{k,-j}{k^3-k\over 12}n.\cr}
$$
Using (3.3), we can rewrite $L_{-1}$ in terms of the $x_k$ and
$\partial\over\partial x_k$'s:
$$L_{-1}={1\over 8n}\sum_{k=1,{\rm odd}}^{2n-1} k(2n-k)x_kx_{2n-k}
         +{1\over 2n}\sum_{k=1,{\rm odd}}^{\infty}(k+2n)x_{k+2n}
         {\partial\over\partial x_k}.\forno$$
We now define in analogy with the untwisted $\hat{sl}_2$ case, i.e.
the KdV hierarchy, the {\it string equation} to be the following
restriction on $\tau\in V_0$:
$$L_{-1}\tau=0.\forno$$
{}From this we deduce (see also [D], [L]) that also
$$\eqalign{L_{-1}\tilde\tau(x,z)=
\{
{1\over 8n}\sum_{k=1,{\rm odd}}^{2n-1}& k(2n-k)(x_k-{2\over kz^k})
         (x_{2n-k}-{2\over (2n-k)z^{2n-k}})\cr
         &+{1\over 2n}\sum_{k=1,{\rm odd}}^{\infty}(k+2n)
         (x_{k+2n}-{2\over (k+2n)z^{k+2n}})
         {\partial\over\partial x_k}
\}\tilde\tau(x,z)=0.\cr}
$$
Now calculating
$$-{\tilde\tau(x,z)L_{-1}\tau(x)\over \tau(x)^2}+
{L_{-1}\tilde\tau(x,z)\over \tau(x)}
$$
explicitly, we deduce that
$$\eqalign{{1\over 2n}\sum_{k=1,{\rm odd}}^{\infty}&(k+2n)x_{k+2n}
  {\partial(\tau(x)^{-1}\tilde\tau(x,z))\over \partial x_k}
  +{1\over 2}z^{-2n}{\tilde\tau(x,z)\over\tau(x)}
  -{1\over 2n}\sum_{k=1,{\rm odd}}^{2n-1} {kx_k\over z^{2n-k}}
  {\tilde\tau(x,z)\over\tau(x)}\cr
  &-{1\over n}{1\over \tau(x)}\sum_{k=1,{\rm odd}}^{\infty} {1\over z^{2n+k}}
  {\partial\tilde\tau(x,z)\over\partial x_k}=0.\cr}
$$
Now compare this with the symbol of $({1\over 2n}ML^{1-2n})_-P$, which is
$$\eqalign{S(({1\over 2n}ML^{1-2n})_-P)=
   -{1\over 2n}\sum_{k=1,{\rm odd}}^{\infty}&(k+2n)x_{k+2n}
   {\partial(\tau(x)^{-1}\tilde\tau(x,z))\over \partial x_k}
   +{1\over 2n}\sum_{k=1,{\rm odd}}^{2n-1} {kx_k\over z^{2n-k}}
   {\tilde\tau(x,z)\over\tau(x)}\cr
   &+{1\over n}{1\over \tau(x)}\sum_{k=1,{\rm odd}}^{\infty} {1\over z^{2n+k}}
   {\partial\tilde\tau(x,z)\over\partial x_k}.\cr}
$$
We thus conclude that the string equation leads to
$$({1\over 2n}ML^{1-2n}-{1\over2}L^{-2n})_-P=0$$
and hence to
$$({1\over 2n}ML^{1-2n}-{1\over2}L^{-2n})_-=0.\forno$$
So
${1\over 2n}ML^{1-2n}-{1\over2}L^{-2n}$
is a differential operator that , moreover, satisfies
$$[L^{2n},{1\over 2n}ML^{1-2n}-{1\over2}L^{-2n}]=1.\forno$$

\noindent \secno.
Notice that since $(L^n)_-=0$ one has
$$({1\over n}ML^{1-n})_-=(({1\over n}ML^{1-n})_-L^n)_-=L^{-n},$$
so also ${1\over n}ML^{1-n}-L^{-n}$ is a differential operator that satisfies
$$[L^n,{1\over n}ML^{1-n}-L^{-n}]=1.$$

\beginsection \chapno . Extra constraints\par
\secno .
{\it From now on we assume that $\tau$ is any solution
of the BKP hierarchy that satisfies:}
$$\eqalign{{\partial\tau\over\partial x_{kn}}&=0\quad\hbox{for}\
k=1,3,5,\ldots\quad\hbox{\rm and}\cr
L_{-1}\tau&=0.\cr}
$$
Hence $(L^n)_-=0$ and
$({1\over 2n}ML^{1-2n}-{1\over2}L^{-2n})_-=0$.
Taking the formal adjoint of these operators one deduces
$(\partial L^n\partial^{-1})_-=0$ and
$({1\over 2n}\partial ML^{1-2n}\partial^{-1}
-{1\over2}\partial L^{-2n}\partial^{-1})_-=0$.
Hence more generally we have for all $p,q\in{\bf Z}_+$:
$$\eqalign{(({1\over 2n}ML^{1-2n}-{1\over2}L^{-2n})^qL^{pn})_-&=0,\cr
(\partial ({1\over 2n} ML^{1-2n}
-{1\over2} L^{-2n})^qL^{pn}\partial^{-1})_-&=0.\cr}
\forno
$$
Now using (4.16) and (4.18) one shows the following
\proclaim Lemma \prono . For all $p,q\in {\bf Z}_+$ one has
$$\eqalign{{\rm Res}_{z=0}dz z^{qn}({1\over 2n}z^{1-n}
{\partial\over\partial z}z^{-n})^p(w(x,z))w^*(x',z)&=0\cr
{\rm Res}_{z=0}dz z^{qn}({1\over 2n}z^{1-n}
{\partial\over\partial z}z^{-n})^p(w^*(x,-z))w(x',-z)&=0\cr}
\forno$$

\proof{ The proof of this lemma is similar to the
proof of lemma 6.1 of [L]}
In terms of the fermionic fields this means
\proclaim Corollary \prono .
For all $p,q\in {\bf Z}_+$ one has
$$\eqalign{{\rm Res}_{z=0}dz z^{qn\over 2}({1\over n}z^{1-n\over 2}
{\partial\over\partial z}z^{1-n\over 2})^p({\phi^+(z)\tau\over\tau})
\otimes\partial({\phi^-(z)\tau\over \tau})&=0\cr
{\rm Res}_{z=0}dz z^{qn\over 2}({1\over n}z^{1-n\over 2}
{\partial\over\partial z}z^{1-n\over 2})^p(\partial({\phi^+(z)\tau\over\tau}))
\otimes{\phi^-(z)\tau\over \tau}&=0\cr}
\forno$$

\noindent\secno .
In the rest of this section the following lemma will be crucial:

\proclaim Lemma \prono .
$$\phi^+(u)\tau\otimes{\partial\over\partial x_1}({\phi^-(v)\tau\over\tau})
  -\phi^-(v)\tau\otimes{\partial\over\partial x_1}({\phi^+(u)\tau\over\tau})
  =-{\rm Res}_{z=0}dz\phi^+(z):\phi^+(u)\phi^-(v):\tau\otimes
   {\partial\over\partial x_1}({\phi^-(z)\tau\over\tau}).$$

\proof{ The bilinear identity (4.19) is equivalent to
$${\rm Res}_{z=0}dz\phi^+(z)\tau\otimes
   {\partial\over\partial x_1}({\phi^-(z)\tau\over\tau})=0.
\forno$$
Now let $2(uv)^{1\over 2}\phi^+(u)\phi^-(v)\otimes 1$
act on this identity, then one obtains:
$$\eqalign{{\rm Res}_{z=0}&{dz\over z}
{1+(v/u)^{1\over 2}\over {1-(v/u)^{1\over 2}}}
{1-(z/u)^{1\over 2}\over {1+(z/u)^{1\over 2}}}
{1+(z/v)^{1\over 2}\over {1-(z/v)^{1\over 2}}}
\exp( -\sum_{k<0} {u^{-k}+z^{-k}-v^{-k}\over k}\alpha_k)\times\cr
&\exp( -\sum_{k>0} {u^{-k}+z^{-k}-v^{-k}\over k}\alpha_k)
\tau\otimes{\partial\over\partial x_1}
({\exp (\sum_{k<0} {z^{-k}\over k}\alpha_k)
   \exp (\sum_{k>0} {z^{-k}\over k}\alpha_k)
   \tau\over\tau})=0.\cr}
\forno$$
Now using the fact that
${1+w\over 1-w}=2\delta(w)-{1+w^{-1}\over 1-w^{-1}}$,
then (7.5) reduces to
$$\eqalign{2(uv)^{1\over 2}&
  (\phi^+(u)\tau\otimes{\partial\over\partial x_1}
  ({\phi^-(v)\tau\over\tau})
  -\phi^-(v)\tau\otimes{\partial\over\partial x_1}({\phi^+(u)\tau\over\tau})
  )\cr
&+
{\rm Res}_{z=0}{dz\over z}
{1+(v/u)^{1\over 2}\over {1-(v/u)^{1\over 2}}}
{1-(u/z)^{1\over 2}\over {1+(u/z)^{1\over 2}}}
{1+(v/z)^{1\over 2}\over {1-(v/z)^{1\over 2}}}
\exp (-\sum_{k<0} {u^{-k}+z^{-k}-v^{-k}\over k}\alpha_k)\times\cr
&\exp (-\sum_{k>0} {u^{-k}+z^{-k}-v^{-k}\over k}\alpha_k)
\tau\otimes{\partial\over\partial x_1}
({\exp (\sum_{k<0} {z^{-k}\over k}\alpha_k)
   \exp (\sum_{k>0} {z^{-k}\over k}\alpha_k)
   \tau\over\tau})=0.\cr}
\forno$$
Now the last term on the left--hand--side is equal to
$$\eqalign{&{\rm Res}_{z=0}dz
2(uv)^{1\over2}\phi^+(z)\phi^+(u)\phi^-(v)\tau\otimes
   {\partial\over\partial x_1}({\phi^-(z)\tau\over\tau})=\cr
&{\rm Res}_{z=0}dz 2(uv)^{1\over2}\phi^+(z):\phi^+(u)\phi^-(v):\tau\otimes
   {\partial\over\partial x_1}({\phi^-(z)\tau\over\tau}).\cr}
$$
}
\noindent\secno .
Define
$$W^{(p+1)}_{{q\over 2}-p}=
{\rm Res}_{z=0}dz z^{qn\over 2}({1\over n}y^{1-n\over 2}
{\partial\over\partial y}y^{1-n\over 2})^p:\phi^+(y)\phi^-(z):|_{y=z},
\forno$$
then from lemma 7.3 and corollary 7.2 we deduce that
$${\rm Res}_{z=0}dz \phi^+(z)W^{(p+1)}_{{q\over 2}-p}\tau
\otimes{\partial\over\partial x_1}({\phi^-(z)\tau\over \tau})
=0,
$$
or explicitly in terms of the $x_k$ and ${\partial\over\partial x_k}$'s:
$${\rm Res}_{z=0}{dz\over z}
e^{x\cdot z}
\exp (-2\sum_{k>0} {z^{-k}\over k}{\partial\over \partial x_k})
(W^{(p+1)}_{{q\over2}-p}\tau(x))
{\partial\over\partial x_1'}
(e^{-x'\cdot z}
{\exp (2\sum_{k>0} {z^{-k}\over k}{\partial\over \partial x_k'})
\tau(x')\over \tau(x')}
)
=0.\forno
$$
First take $x_k=x_k'$ for all $k=1,3,\ldots$, then one deduces that
$$
{\partial\over\partial x_1}({W^{(p+1)}_{{q\over2}-p}\tau(x)\over \tau(x)})
=0.\forno
$$
Now divide (7.8) by $\tau(x)$, then
$${\rm Res}_{z=0}{dz}
w(x,z)\exp (-2\sum_{k>0} {z^{-k}\over k}{\partial\over \partial x_k})
({W^{(p+1)}_{{q\over2}-p}\tau(x)\over \tau(x)})w^*(x',z)=0.\forno
$$
Now subtract a multiple of the bilinear identity (4.19), then one obtains
$${\rm Res}_{z=0}{dz}
w(x,z)(\exp (-2\sum_{k>0} {z^{-k}\over k}{\partial\over \partial x_k})-1)
({W^{(p+1)}_{{q\over2}-p}\tau(x)\over \tau(x)})w^*(x',z)=0.\forno
$$
Define
$$S_{pq}=S_{pq}(x,z)=
  (\exp (-2\sum_{k>0} {z^{-k}\over k}{\partial\over \partial x_k})-1)
({W^{(p+1)}_{{q\over2}-p}\tau(x)\over \tau(x)}),
$$
then (7.9) implies that $\partial\circ
S_{pq}(x,\partial)=S_{pq}(x,\partial)\circ
\partial$. Using this and lemma 4.1, we deduce that
$$(PS_{pq}(\partial P\partial^{-1})^*)_-=(PS_{pq}P^{-1})_-
=PS_{pq}P^{-1}=0.
$$
So $S_{pq}(x,z)=0$ and hence
$${W^{(p+1)}_{{q\over2}-p}\tau(x)\over \tau(x)}=\hbox{\rm constant}
\quad\hbox{\rm for}\ p,q\in {\bf Z}_+.
\forno$$
In the next section we will see that the
$W^{(p+1)}_{{q\over2}-p}$ form a subalgebra of $W_{1+\infty}$.

\beginsection \chapno . The $BW_{1+\infty}$ constraints\par
\secno .
The Lie algebras $\gli, \oi, \glio$ and $\oio$ all have a natural action
on the space of column vectors, viz.,
let ${\bf C}^{\infty}=\bigoplus_{k\in{\bf Z}} e_k$, then
$E_{ij}e_k=\delta_{jk}e_i$. By identifying $e_k$ with $s^{-k}$, we
can embed the algebra ${\bf D}$ of differential operators on the circle,
with basis $-s^{j+k}({\partial\over \partial s})^k$ ($j\in{\bf Z},
k\in{\bf Z}_+$), in $\glio$:
$$\eqalign{\rho:{\bf D}&\to\glio\cr
           \rho(-s^{j+k}({\partial\over \partial s})^k)&=
 \sum_{m\in{\bf Z}}-m(m-1)\cdots(m-k+1)E_{-m-j,-m}\cr}\forno
$$
It is straightforward to check that the 2--cocycle
$\mu$ on $\glio$ induces the following 2--cocycle on ${\bf D}$:
$$\mu(-s^{i+j}({\partial\over\partial s})^j,
      -s^{k+\ell}({\partial\over\partial s})^{\ell})=
\delta_{i,-k}(-)^j j!\ell !{i+j\choose j+\ell+1}.\forno$$
This cocycle was discovered by Kac and Peterson in [KP]
(see also [R], [KR]). In this way we have defined a central extension of ${\bf
D}$,
which we denote by $W_{1+\infty}={\bf D}\oplus {\bf C}c_A$,
the Lie bracket on $W_{1+\infty}$ is given by
$$\eqalign{[-s^{i+j}&({\partial\over\partial s})^j+\alpha c_A,
      -s^{k+\ell}({\partial\over\partial s})^{\ell}+\beta c_A]=\cr
&\sum_{m=0}^{\max (j,\ell)}m!({i+j\choose m}{\ell\choose m}
-{k+\ell\choose m}{j\choose m})
(-s^{i+j+k+\ell-m}({\partial\over\partial s})^{j+\ell-m})+
\delta_{i,-k}(-)^j j!\ell !{i+j\choose j+\ell+1}c_A.\cr}\forno$$

Let $D_m=s^m{\partial\over\partial s^m}$ and set $D=D_1$, then we can rewrite
the elements $-s^{i+j}({\partial\over\partial s})^j$, viz. ,
$$-s^{i+j}({\partial\over\partial s})^j
=-s^{i}D(D-1)(D-2)\cdots(D-j+1).\forno$$
Then for $k\ge 0$ the 2--cocycle is as follows [KR]:
$$\mu (s^kf(D), s^{\ell}g(D))=
\cases{\sum_{-k\le j\le -1}f(j)g(j+k)& if $k=-\ell$\cr
       0& otherwise\cr}\forno$$

\noindent\secno .
Now replace $s$ by $t^{1\over 2}$ and write
$2t^{1\over 2}{\partial\over\partial t}$
instead of $\partial\over\partial t^{1\over 2}$. Then a new basis
of ${\bf D}$ is given by $-t^{{j\over 2}+k}({\partial\over\partial t})^k$.
It is then straightforward to check that the anti--involution $\iota$
defined by (2.4) induces
$$\iota(t^{1\over 2})=-t^{1\over 2},\quad  \iota(t^{1\over2}{\partial\over
\partial t}t^{-{1\over2}})
=-t^{1\over2}{\partial\over \partial t}t^{-{1\over2}}\quad\hbox{\rm
and}\quad\iota(D)=-D.
\forno
$$
Hence, it induces the following anti--involution on ${\bf D}$:
$$\iota(t^{1\over2}t^{{j\over 2}+k}
({\partial\over \partial t})^kt^{-{1\over2}})=
(-)^{j+k}t^{1\over2}({\partial\over \partial t})^k
t^{{j\over 2}+k}t^{-{1\over2}}.
\forno$$
Define ${\bf D}^B$=${\bf D}\cap\oio=\{w\in {\bf D}|\iota(w)=-w\}$, it is
spanned by
the elements
$$\eqalign{w_{j\over2}^{(k+1)}&=
-t^{1\over2}(t^{{j\over 2}+k}
({\partial\over \partial t})^k-
(-)^{j+k}({\partial\over \partial t})^k
t^{{j\over 2}+k})t^{-{1\over2}}\cr
&=-t^{1\over2}(t^{{j\over 2}+k}
({\partial\over \partial t})^k-
(-)^{j+k}\sum_{\ell=0}^k {k\choose \ell}\ell !
{{j\over 2}+k\choose\ell}t^{{j\over 2}+k-\ell}
({\partial\over \partial t})^{k-\ell})t^{-{1\over2}}.\cr}
\forno$$
The restriction of the 2--cocycle $\mu$ on ${\bf D}$, given
by (8.2), induces a 2--cocycle on ${\bf D}^B$, which we shall not calculate
explicitly here. It defines a central extension
$BW_{1+\infty}={\bf D}^B\oplus{\bf C}c_B$ of ${\bf D}^B$, with Lie bracket
$$[a+\alpha c_B,b+\beta c_B]=ab-ba+{1\over 2}\mu(a,b)c_B,$$
for $a,b\in {\bf D}^B$ and $\alpha,\beta\in{\bf C}$.

\noindent\secno .
We work out
$$\eqalign{:{\partial^p\phi^+(z)\over\partial z^p}\phi^-(z):&=
\sum_{k,\ell\in{\bf Z}}-p!{\ell-{1\over 2}\choose p}\hat\pi(F_{k+\ell,-\ell})
z^{-{k\over 2}-p-1}\cr
&=
\sum_{k,\ell\in{\bf Z}}-p!{\ell-{1\over 2}\choose p}
(\hat\pi(E_{-k-\ell,-\ell})-(-)^k\hat\pi(E_{\ell,k+\ell}))
z^{-{k\over 2}-p-1}\cr&=
\sum_{k,\ell\in{\bf Z}}-p!
({\ell-{1\over 2}\choose p}-(-)^k{-k-\ell-{1\over2}\choose p})
\hat\pi(E_{-k-\ell,-\ell})z^{-{k\over 2}-p-1}\cr&=
\sum_{k,\ell\in{\bf Z}}-p!
({\ell-{1\over 2}\choose p}-(-)^{k+p}{k+\ell+p-{1\over2}\choose p})
\hat\pi(E_{-k-\ell,-\ell})z^{-{k\over 2}-p-1}\cr&=
\hat\pi(-t^{1\over2}(t^{{k\over 2}+p}
({\partial\over \partial t})^p-
(-)^{k+p}({\partial\over \partial t})^p
t^{{k\over 2}+p})t^{-{1\over2}})z^{-{k\over 2}-p-1}\cr&=
\hat\pi(w_{k\over2}^{(p+1)})z^{-{k\over 2}-p-1}.\cr}
\forno
$$
\noindent\secno .
We want to calculate $W^{(p+1)}_{{q\over2}-p}$.
For this purpose we write
$$(z^{1-n\over 2}{\partial\over\partial z}z^{1-n\over 2})^p=
\sum_{\ell=0}^p c(\ell,p)z^{-np+\ell}({\partial\over\partial z})^{\ell}.
$$
Then
$$\eqalign{W^{(p+1)}_{{q\over2}-p}&=
{\rm Res}_{z=0}dz z^{qn\over 2}({1\over n}y^{1-n\over 2}
{\partial\over\partial y}y^{1-n\over 2})^p:\phi^+(y)\phi^-(z):|_{y=z}\cr
&={\rm Res}_{z=0}dz {1\over n^p}
\sum_{\ell=0}^p c(\ell,p)z^{{qn\over 2}-np+\ell}
:{\partial^{\ell}\phi^+(z)\over\partial z^{\ell}}\phi^-(z):\cr
&={1\over n^p}\sum_{\ell=0}^p c(\ell,p)
\hat\pi(w^{(\ell+1)}_{{qn\over 2}-pn})\cr
&=\hat\pi({1\over n^p}
\sum_{\ell=0}^p c(\ell,p)(-t^{1\over2}(t^{{qn\over 2}-np+\ell}
({\partial\over\partial t})^{\ell})t^{-{1\over2}}
-\iota(-t^{1\over2}(t^{{qn\over 2}-np+\ell}
({\partial\over\partial t})^{\ell})t^{-{1\over2}})))\cr
&=\hat\pi({1\over n^p}(-t^{1\over2}(t^{{qn\over 2}}
(t^{1-n\over 2}{\partial\over\partial t}t^{1-n\over 2})^p)t^{-{1\over2}})
-\iota({1\over n^p}(-t^{1\over2}(t^{{qn\over 2}}
(t^{1-n\over 2}{\partial\over\partial t}t^{1-n\over 2})^p)
t^{-{1\over2}})))\cr
&=\sum_{k\in{\bf Z}}\hat\pi(-t^{1\over2}({1\over n^p}(t^{{qn\over 2}}
(t^{1-n\over 2}{\partial\over\partial t}t^{1-n\over 2})^p
-(-)^{p+q}{1\over n^p}
(t^{1-n\over 2}{\partial\over\partial t}t^{1-n\over 2})^p
t^{{qn\over 2}})
t^{-{1\over2}}))\cr
&=\sum_{k\in{\bf Z}}\hat\pi(-\lambda^{1\over2}(\lambda^{q\over 2}
({\partial\over\partial\lambda})^p-(-)^{p+q}
({\partial\over\partial\lambda})^p\lambda^{q\over 2})
\lambda^{-{1\over 2}}),\cr}
\forno$$
where $\lambda=t^n=s^{2n}$. Hence, from this it is obvious that the elements
$W^{(p+1)}_{{k\over 2}}$, together with $c_B$  spann a $BW_{1+\infty}$--algebra
with $c_B=nI$.

\beginsection \chapno . The calculation of the constants\par
\noindent\secno .
In order to determine the constants on the right--hand--side of (7.12),
we notice that
$$\eqalign{
0&=[W_{-1}^{(2)},-{1\over q+2}W^{(p+1)}_{{q\over 2}-p+1}]\tau\cr
 &=(W^{(p+1)}_{{q\over 2}-p}
 +{1\over 2}\mu(W_{-1}^{(2)},-{1\over q+2}W^{(p+1)}_{{q\over 2}-p+1})
    )\tau.\cr}
\forno$$
It is clear that the cocycle term of (9.1) is 0, except when $q=2p$.

Now
$$W_{-1}^{(2)}=
-2\lambda^{1\over2}{\partial\over\partial\lambda}\lambda^{-{1\over 2}}=
s^{-2n}-{1\over n}s^{1-2n}{\partial\over\partial s}= -{s^{-2n}\over n}(D-n)$$
and
$$\eqalign{&-{1\over 2p+2}W_1^{(p+1)}\cr
&={1\over 2p+2}[s^{3n}D_{2n}(D_{2n}-1)\cdots(D_{2n}-p+1)s^{-n}
-\iota(s^{3n}D_{2n}(D_{2n}-1)\cdots(D_{2n}-p+1)s^{-n})]\cr
&={1\over 2p+2}[s^{2n}(D_{2n}-{1\over 2})(D_{2n}-{3\over
2})\cdots(D_{2n}-p+{1\over 2})-\iota(s^{2n}(D_{2n}-{1\over 2})(D_{2n}-{3\over
2})\cdots(D_{2n}-p+{1\over 2}))]\cr
&={1\over 2p+2}{s^{2n}\over (2n)^p}[(D-n)(D-3n)\cdots (D-(2p-1)n)- (-)^p
  (D+(2p+1)n)(D+(2p-1)n)\cdots (D+3n)].\cr}
$$

Then
using (8.5) we deduce that
$$\eqalign{&\mu(W_{-1}^{(2)},-{1\over 2p+2}W^{(p+1)}_{1})\cr
&={1\over p+1}({1\over 2n})^{p+1}\!\!\!\sum_{-2n\le j\le-1}\!\!
[(j+n)(j-n)\cdots(j-(2p-1)n)-(-)^p
 (j+(2p+1)n)(j+(2p-1)n)\cdots(j+n)]\cr
&={1\over p+1}(\sum_{-{1\over 2}\le k\in {1\over 2n}{\bf Z}<{1\over 2}}+
\sum_{-{1\over 2}< k\in {1\over 2n}{\bf Z}\le{1\over 2}}) \ k(k-1)(k-2)\cdots
(k-p).
\cr}\forno
$$
Hence we can state the main Theorem of this paper:
\proclaim Theorem \prono . The following two constraints
on a BKP $\tau$--function are equivalent:
$${\partial\tau\over\partial x_{jn}}=0\qquad\hbox{for}\
j=1,3,5,\ldots\quad\hbox{\rm and}\leqno{(1):}$$
$$\{\sum_{k=1,{\rm odd}}^{2n-1} k(2n-k)x_kx_{2n-k}+4\sum_{k=1,{\rm
odd}}^{\infty}
(2n+k)x_{2n+k}{\partial\over\partial x_k}\}\tau=0.$$
$$\{W^{(p+1)}_{{q\over2}-p}+{\delta_{2q,p}\over 2}c_{p+1}\}\tau=0,\leqno{(2):}
$$
for $p,q\in{\bf Z}_+$, where
$$c_{p+1}={1\over p+1}(\sum_{-{1\over 2}\le k\in {1\over 2n}{\bf Z}<{1\over
2}}+
\sum_{-{1\over 2}< k\in {1\over 2n}{\bf Z}\le{1\over 2}}) \ k(k-1)(k-2)\cdots
(k-p).
$$
\par
\proof{ The proof of this theorem is now obvious, since (2) clearly implies
(1)}

Notice that $c_1=0$, $c_2={2n^2+1\over 6n}=8({1\over 16n}+ {n^2-1\over 24n})$.

\beginsection \chapno. Appendix

In this appendix we show that it is possible to express
$$\hat W^{(p+1)}_{{q\over2}-p}=W^{(p+1)}_{{q\over2}-p}+{\delta_{2q,p}\over
2}c_{p+1}$$
in terms of the
$\alpha_j$'s.
Recall (7.7):
$$\eqalign{W^{(p+1)}_{{q\over 2}-p}&=
{\rm Res}_{z=0}dz z^{qn\over 2}({1\over n}y^{1-n\over 2}
{\partial\over\partial y}y^{1-n\over 2})^p:\phi^+(y)\phi^-(z):|_{y=z}\cr
&={\rm Res}_{z=0}dz z^{(q+1)n-1\over 2}
({\partial\over\partial y^n})^p y^{1-n\over 2}:\phi^+(y)\phi^-(z):|_{y=z}\cr}
\forno$$
Using (3.2), it is straightforward to check that
$$(y^n-z^n) y^{1-n\over 2}:\phi^+(y)\phi^-(z):=
{1\over 2}(\sum_{k=0}^{2n-1} +\sum_{k=1}^{2n})y^{n-k\over 2}z^{k-1\over 2}
(X(y,z)-1),\forno$$
where
$$X(y,z)=\exp (-\sum_{k<0} {y^{-k}-z^{-k}\over k}\alpha_k)
         \exp (-\sum_{k>0} {y^{-k}-z^{-k}\over k}\alpha_k).\forno$$
Hence
$$\eqalign{W^{(p+1)}_{{q\over 2}-p}&=
{\rm Res}_{z=0}{dz\over z} {1\over 2p+2}({\partial\over\partial
y^n})^{p+1}\{(\sum_{k=0}^{2n-1} +\sum_{k=1}^{2n})
(y^{n})^{n-k\over 2n}(z^n)^{(q+1)n+k\over 2n}
(X(y,z)-1)|_{y=z}\}\cr
&={\rm Res}_{z=0}{dz\over z} {1\over 2p+2}
\sum_{\ell=0}^{p} {p+1\choose \ell}
({\partial\over\partial y})^{\ell}\{(\sum_{k=0}^{2n-1} +\sum_{k=1}^{2n})
y^{n-k\over 2n}z^{(q+1)n+k\over 2n}\}
{\partial^{p+1-\ell}X(y^{1\over n},z^{1\over n})
\over\partial y^{p+1-\ell}}|_{y=z}\cr
&={\rm Res}_{z=0}{dz\over z} {1\over 2p+2}
\sum_{\ell=0}^{p} {p+1\choose \ell}\ell !
(\sum_{k=0}^{2n-1} +\sum_{k=1}^{2n}){{n-k\over 2n}\choose \ell}z^{{q\over
2}+1-\ell}{\partial^{p+1-\ell}X(y^{1\over n},z^{1\over n})
\over\partial y^{p+1-\ell}}|_{y=z}.\cr}
\forno
$$
Notice that $W_{q\over2}^{(p+1)}=w_{q\over2}^{(p+1)}$ for
$n=1$.

Since
$$c_{\ell}=
(\ell -1)!
(\sum_{k=0}^{2n-1} +\sum_{k=1}^{2n}){{n-k\over 2n}\choose \ell},$$
one finds that

$$\eqalign{\hat W^{(p+1)}_{{q\over 2}-p}&=
{\rm Res}_{z=0}{dz\over z} {1\over 2p+2}({\partial\over\partial
y^n})^{p+1}\{(\sum_{k=0}^{2n-1} +\sum_{k=1}^{2n})
(y^{n})^{n-k\over 2n}(z^n)^{(q+1)n+k\over 2n}
X(y,z)|_{y=z}\}\cr
&={\rm Res}_{z=0}{dz} {z^{q\over 2}\over 2p+2}(4n{\partial^{p+1}X(y^{1\over
n},z^{1\over n})
\over\partial y^{p+1}}+
\sum_{\ell=2}^{p+1} {p+1\choose \ell}\ell c_{\ell}
z^{-\ell}{\partial^{p+1-\ell}X(y^{1\over n},z^{1\over n})
\over\partial y^{p+1-\ell}})|_{y=z}.
\cr}
\forno
$$
The right--hand--side of (10.5) is some expression in the $\alpha_k$'s,
here are a few of the fields ${\partial^{m}X(y^{1\over n},z^{1\over n})
\over\partial y^{m}})|_{y=z}$:
$$\eqalign{{\partial X(y^{1\over n},z^{1\over n})
\over\partial y})|_{y=z}&={1\over n}\sum_k \alpha_k z^{-{k+n\over n}}\cr
{\partial^{2}X(y^{1\over n},z^{1\over n})
\over\partial y^{2}})|_{y=z}&={1\over n^2}\sum_k ( \alpha^{(2)}_k
-(k+n)\alpha_k )z^{-{k+2n\over n}}\cr
{\partial^{3}X(y^{1\over n},z^{1\over n})
\over\partial y^{3}})|_{y=z}&={1\over n^3}\sum_k ( \alpha^{(3)}_k -{3\over
2}(k+2n) \alpha^{(2)}_k+(k+n)(k+2n)\alpha_k  )z^{-{k+3n\over n}},\cr}
$$
where $\alpha_k^{(2)}=\sum_j :\alpha_{-j}\alpha_{k+j}:$ and
$\alpha_k^{(3)}=\sum_{i,j} :\alpha_{-i}\alpha_{-j}\alpha_{i+j+k}:$.

\vskip .3cm
\noindent{\bf References} \vskip .3cm
\halign{#\hfil&\quad\vtop{\parindent=0pt\hsize=41em\strut#\strut}\hfill\cr

[{\bf AV} ]&{M. Adler and P. van Moerbeke, A Matrix Integral Solution to
Two--dimensional $W_p$--Gravity, Comm. Math. Phys. {\bf  147} (1992),
25--56.}\cr

[{\bf DJKM1} ]&{E. Date, M. Jimbo, M. Kashiwara and T. Miwa,
Transformation groups for soliton equations.  Euclidean Lie
algebras and reduction of the KP hierarchy , Publ. Res. Inst.
Math. Sci. {\bf  18} (1982),  1077--1110.}\cr

[{\bf DJKM2} ]&{E. Date, M. Jimbo, M. Kashiwara and T. Miwa,
Transformation groups for soliton equations , in:
Nonlinear integrable systems---classical theory and quantum theory
eds M. Jimbo and T. Miwa, World Scientific, 1983),  39--120.}\cr

[{\bf DJKM3} ]&{E. Date, M. Jimbo, M. Kashiwara and T. Miwa,
Transformation groups for soliton equations IV. A new hierarchy of
soliton equations of KP type, Physica 4D (1982), 343--365.}\cr

[{\bf D} ]&{L.A. Dickey , Additional symmetries of KP, Grassmannian, and the
string equation II, preprint University of Oklahoma  (1992).}\cr

[{\bf Dij} ]&{R. Dijkgraaf , Intersection Theory, Integrable Hierarchies and
Topological Field Theory, preprint IASSNS--HEP--91, hep-th 9201003.}\cr

[{\bf FKN}]&{M. Fukuma, H. Kawai and R. Nakayama,
Infinite Dimensional Grassmannian Structure of Two--Dimensional Quantum
Gravity, Comm. Math. Phys.{\bf  143} (1992),  371--403.}\cr

[{\bf G} ]&{J. Goeree, $W$--cinstraints in 2d quantum gravity, Nucl. Phys.
{\bf  B358} (  1991),  737--157.}\cr

[{\bf H} ]&{P. van den Heuvel, Polynomial solutions of hierarchies of
differential equations, Masters's thesis Univ. Utrecht (1992).}\cr

[{\bf K} ]&{V.G. Kac , Infinite dimensional Lie algebras,
Progress in Math., vol. 44, Brikh\"{a}user, Boston, 1983; 2nd
ed., Cambridge Univ. Press, 1985; 3d ed., Cambridge Univ. Press,
1990.}\cr

[{\bf KL}]&{V. Kac and J. van de Leur, The $n$--Component KP hierarchy and
Representation Theory., in Important Developments in Soliton Theory, eds. A.S.
Fokas and V.E. Zakharov. Springer Series in Nonlinear Dynamics, (1993),
302--343.}\cr

[{\bf KP} ]&{V.G. Kac and D.H. Peterson , Spin and wedge
representations of infinite dimensional Lie algebras and groups
, Proc. Nat. Acad. Sci U.S.A. (1981),  3308--3312.
}\cr

[{\bf KR}]&{V. Kac and A. Radul, Quasifinite highest weight
modules over the Lie algebra of differential operators on the circle
, Comm. Math. Phys. {\bf  157 } (1993),  429-457.
}\cr

[{\bf tKL} ]&{F. ten Kroode and J. van de Leur,
Level one representations of the
twisted affine algebras $A_n^{(2)}$ and $D_n^{(2)}$, Acta Appl. Math.
{\bf 27} (1992), 153 -- 224.
}\cr

[{\bf L} ]&{J. van de Leur, KdV type hierarchies, the string equation
and $W_{1+\infty}$ constraints, preprint Univ. Utrecht and hep-th
9403080, to appear in Journal of Geometry and Physics.}\cr

[{\bf R} ]&{A.O. Radul, Lie algebras of differential operators,
their central extensions, and W--algebras , Funct. Anal. and its Appl.
{\bf  25
} (1991),  33--49.}\cr

[{\bf Sh1} ]&{T. Shiota, Characterization of Jacobian
varieties in terms of soliton equations , Invent. Math. {\bf  83
} (1986),  333--382.}\cr

[{\bf Sh2} ]&{T. Shiota, Prym varieties and soliton equations, in:
Infinite dimensional Lie algebras and groups, ed. V.G. Kac,
Adv, Ser. in
Math. phys. 7, world Sci., 1989, 407--448.}\cr

[{\bf Y} ]&{Y.--C. You,Polynomial solutions of the BKP hierarchy
and projective representations of  symmetric groups, in:
Infinite dimensional Lie algebras and groups, ed. V.G. Kac,
Adv, Ser. in
Math. phys. 7, world Sci., 1989, 449--466.}\cr}

\vskip .5cm
\noindent
{\bf JOHAN VAN DE LEUR}

\noindent FACULTY OF APPLIED MATHEMATICS

\noindent UNIVERSITY OF TWENTE

\noindent P.O. BOX 217

\noindent 7500 AE ENSCHEDE

\noindent THE NETHERLANDS

\end